\documentclass[twocolumn,showpacs,preprintnumbers,amsmath,amssymb,floatfix]{revtex4-1}
\usepackage{graphicx}
\usepackage{epsfig}
\usepackage{dcolumn}
\usepackage{bm}

\begin{document}

\title{Anisotropic $H_{c2}$ determined up to 92 T and the signature of multi-band superconductivity in Ca$_{10}$(Pt$_{4}$As$_{8}$)((Fe$_{1-x}$Pt$_{x}$)$_{2}$As$_{2}$)$_{5}$ superconductor}

\author{Eundeok Mun$^{1}$, Ni Ni$^{2}$, Jared M. Allred$^{2}$, Robert J. Cava$^{2}$, Oscar Ayala$^{1}$, Ross D. McDonald$^{1}$, Neil Harrison$^{1}$, Vivien S. Zapf$^{1}$}
\affiliation{$^{1}$National High Magnetic Field Laboratory, Los Alamos National Laboratory, Los Alamos, New Mexico 87545, USA}%
\affiliation{$^{2}$Department of Chemistry, Princeton University, Princeton, NJ 08544, USA}

\begin{abstract}
The upper critical fields, $H_{c2}$($T$), of single crystals of the superconductor Ca$_{10}$(Pt$_{4-\delta}$As$_{8}$)((Fe$_{0.97}$Pt$_{0.03}$)$_{2}$As$_{2}$)$_{5}$ ($\delta$ $\approx$
0.246) are determined over a wide range of temperatures down to $T$ = 1.42 K and magnetic fields of up to $\mu_{0}H$ $\simeq$ 92 T. The measurements of anisotropic $H_{c2}$($T$) curves
are performed in pulsed magnetic fields using radio-frequency contactless penetration depth measurements for magnetic field applied both parallel and perpendicular to the
\textbf{ab}-plane. Whereas a clear upward curvature in $H_{c2}^{\parallel\textbf{c}}$($T$) along \textbf{H}$\parallel$\textbf{c} is observed with decreasing temperature, the
$H_{c2}^{\parallel\textbf{ab}}$($T$) along \textbf{H}$\parallel$\textbf{ab} shows a flattening at low temperatures. The rapid increase of the $H_{c2}^{\parallel\textbf{c}}$($T$) at
low temperatures suggests that the superconductivity can be described by two dominating bands. The anisotropy parameter, $\gamma_{H}$ $\equiv$
$H_{c2}^{\parallel\textbf{ab}}$/$H_{c2}^{\parallel\textbf{c}}$, is $\sim$7 close to $T_{c}$ and decreases considerably to $\sim$1 with decreasing temperature, showing rather weak
anisotropy at low temperatures.
\end{abstract}


\maketitle

After the discovery of superconductivity in the LaFeAsO$_{1-x}$F$_{x}$ system, enormous research efforts have been devoted to searching for new Fe-based superconductors with higher
transition temperatures, $T_{c}$, and they have been discovered and categorized into several types \cite{Kamihara2008, Hsu2008, Rotter2008, Pitcher2008, Parker2008, Guo2010}. All of
these materials have layers with edge-sharing tetrahedra, where Fe atoms are surrounded by four As or Se atoms. Recently Ca$_{10}$(Pt$_{n}$As$_{8}$)(Fe$_{2}$As$_{2}$)$_{5}$ systems
have shown that the $T_{c}$ can be as high as $\sim$38 K by substituting Pt for Fe in the Fe$_{2}$As$_{2}$ layer \cite{Ni2011, Lohnert, Kakiya}. These new structures have stacks of
Ca-(Pt$_{n}$As$_{8}$)-Ca-(Fe$_{2}$As$_{2}$), forming a triclinic $P\overline{1}$, 10-3-8 phase with $n$ = 3 and tetragonal $P4/n$, 10-4-8 phase with $n$ = 4 \cite{Ni2011}.

Upper critical field, $H_{c2}$, measurements of many Fe-based superconductors have revealed similar temperature dependences of the anisotropy ratio, $\gamma_{H}$ $\equiv$
$H_{c2}^{\parallel\textbf{ab}}$/$H_{c2}^{\parallel\textbf{c}}$, decreasing with decreasing temperature, with $\gamma_{H}\lesssim 10$ close to the $T_{c}$ \cite{Hunte2008,
Jaroszynski2008, Lee2008, Altarawneh2008, Ni2008, Kano2009, Khim2010, Mun2011}. As evidenced by both experiments and band-structure calculations, the Fe-based superconductors are a
multi-band system \cite{Hunte2008, Jaroszynski2008, Kano2009, Lee2008, Eschrig2009, Weyeneth2009, Paglione2010}. Multi-band superconductivity can be manifested in the temperature
dependence of the anisotropic parameters. For example, $\gamma_{H}$ of MgB$_{2}$ decreases with increasing temperature \cite{Gurevich2003}, whereas for many Fe-based materials
$\gamma_{H}$ increases on warming \cite{Hunte2008, Kano2009}. For one-band isotropic materials the anisotropic parameters generally have no temperature dependence.

In light of the multi-band superconductivity and the very high $H_{c2}(0)$ of Fe-based superconductors, rapidly exceeding the limit of the commonly available magnetic fields, finding
the true behavior of the $H_{c2}(T)$ and $\gamma_{H}$ down to very low temperatures and high magnetic fields are of great interest. In this communication we present $H_{c2}(T)$ curves
determined from radiofrequency (rf) contactless penetration depth measurements in pulsed magnetic fields up to 92 T, for single crystals of the 10-4-8 phase materials. The high
magnetic fields up to 92 T and temperatures down to 1.42 K, used in this study, enable access to the complete evolution of the $H_{c2}(T)$ curves. We find that as the temperature
decreases, the $H_{c2}^{\parallel\textbf{ab}}(T)$ curve flattens, whereas $H_{c2}^{\parallel\textbf{c}}(T)$ increases steeply, which is shown to be consistent with the multi-band
effects.

\begin{figure}
\centering
\includegraphics[width=0.9\linewidth]{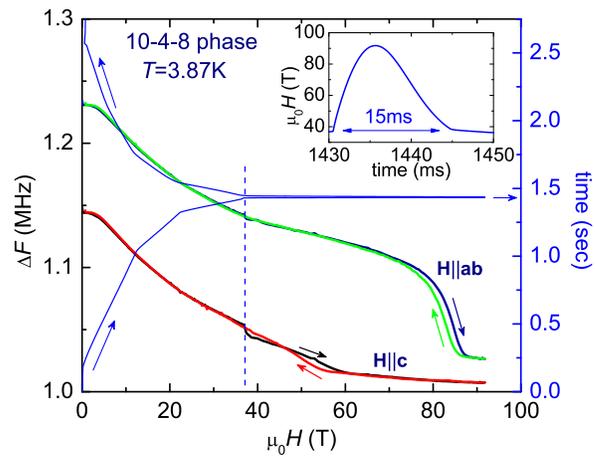}
\caption{(Color online) The representative data sets taken in a multi-shot magnet. Left axis: The rf frequency shift, $\Delta F$, as a function of magnetic field for
\textbf{H}$\parallel$\textbf{ab} and \textbf{H}$\parallel$\textbf{c} at $T$ = 3.87 K. Right axis: The waveform of pulse for the 92 T. The fast sweep rate of magnetic fields up to 92 T
(within 15 ms, see inset) is provided by a capacitor bank on the peak of the slowly sweeping $\sim$ 36 T magnetic field ($\sim$2.5 sec) derived by a motor generator. Because of the
radical change of magnetic field at $\sim$36 T, $\Delta F$ curves show a slope break, as indicated by the vertical dashed line.}
\label{Fig1}%
\end{figure}%

The Ca$_{10}$(Pt$_{4-\delta}$As$_{8}$)((Fe$_{0.97}$Pt$_{0.03}$)$_{2}$As$_{2}$)$_{5}$ ($\delta$ $\approx$ 0.246) samples with $T_{c}$ = 26.5 K, were prepared by heating precursor
materials above 1100 $^{\texttt{o}}$C. The shape of the as-grown crystal is plate-like in the \textbf{ab}-plane with shiny surfaces. Details of sample growth and characterization are
given in Ref. \cite{Ni2011}. To study the anisotropy of $H_{c2}$ up to $\mu_{0}H$ = 92 T, the magnetic field dependence of rf contactless penetration depth, known as proximity
detection oscillator (PDO) \cite{Altarawneh2009}, was measured for magnetic field applied both parallel (\textbf{H}$\parallel$\textbf{ab}) and perpendicular
(\textbf{H}$\parallel$\textbf{c}) to the \textbf{ab}-plane of the tetragonal structure in two different pulsed magnet systems. The rf technique has been shown to be a sensitive and
accurate method for determining the $H_{c2}$ of superconductors, especially in pulsed magnetic fields \cite{Mielke2001}. Details about this technique can be found in Refs.
\cite{Coffey2000, Mielke2001, Altarawneh2008, Altarawneh2009, Mun2011}. For magnetic fields up to 60 T, a short-pulse magnet was used with a 10 ms rising and 40 ms falling time.
Measurements were extended to 92 T in the 95 T multi-shot magnet system at the National High Magnetic Field Laboratory Pulsed Field Facility: Here the magnetic fields up to $\sim$36 T
are provided by a motor-generator-driven outsert magnet with a slower sweep rate and longer pulse length ($\sim$2.5 s) and the remaining magnetic field from $\sim$36 T up to 92 T is
provided (within 15 ms) by a capacitor-bank-driven insert magnet. The magnetic field versus time profile of the insert magnet is shown in the inset to Fig. \ref{Fig1}. Examples of
resonance frequency shift ($\Delta$$F$) measured using the rf technique are shown in Fig. \ref{Fig1}, where changes in the resistivity and magnetic susceptibility drive a shift in the
resonance frequency of an oscillator circuit. At $\sim$36 T, $\Delta$$F$ shows a feature due to the rapid shift in the rate of change of magnetic field, as indicated by the vertical
line in Fig. \ref{Fig1}. The magnetic field was determined using a pick-up coil and calibrated by measuring quantum oscillations in copper. To minimize the eddy current heating and
dissipative vortex motion due to the fast sweep rate of pulsed magnetic fields, small single crystals were chosen, typically $\sim$ 0.2$\times$0.2$\times$0.05 mm$^{3}$. Note that in
order to reduce open-loop area of samples for \textbf{H}$\parallel$\textbf{c}, a smaller size sample ($\sim$ 0.12$\times$0.12$\times$0.05 mm$^{3}$) was used for measurements up to 92
T in the multi-shot magnet. No significant difference between 60 T and 92 T measurements was detected.

\begin{figure}
\centering
\includegraphics[width=0.95\linewidth]{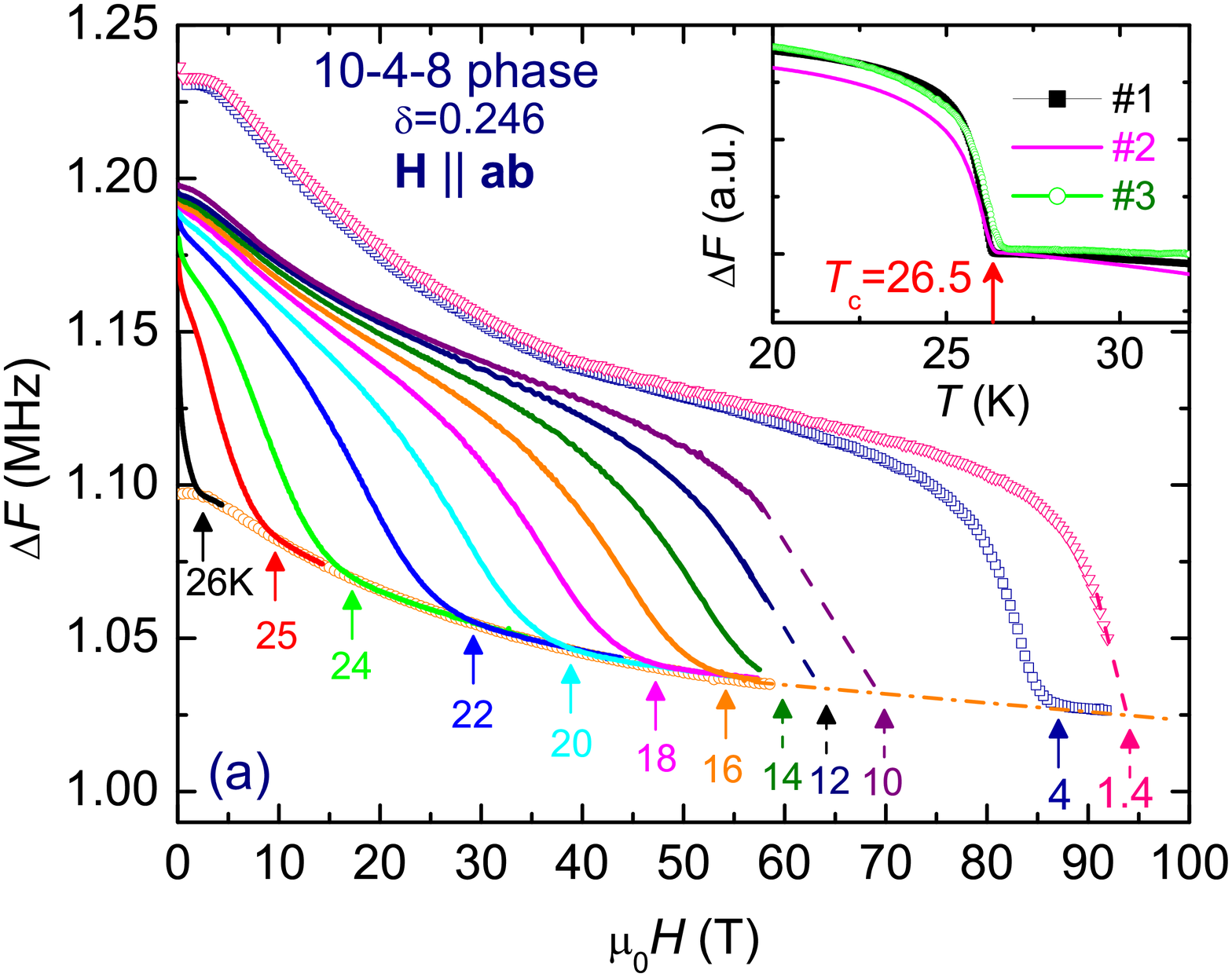}
\includegraphics[width=0.95\linewidth]{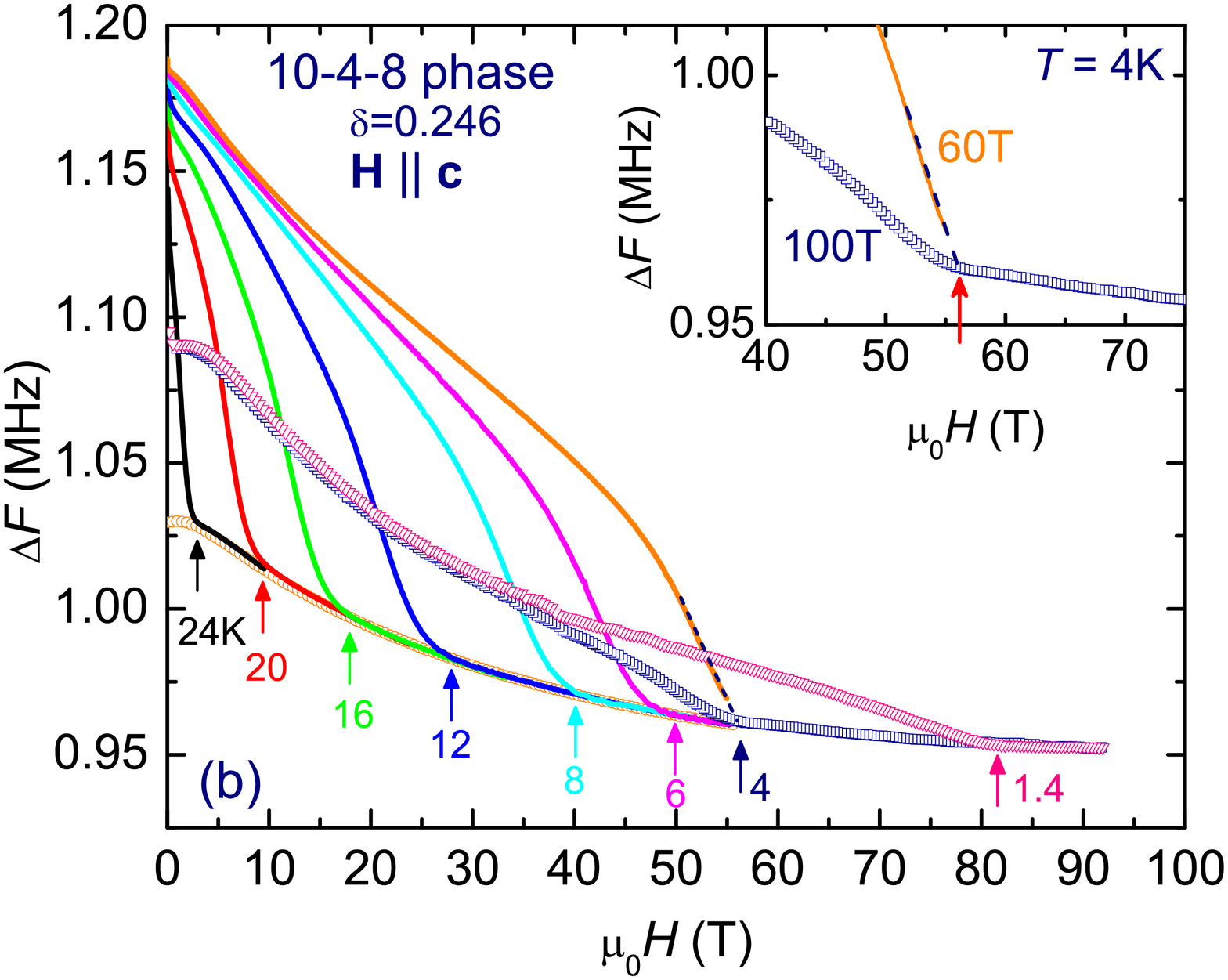}
\caption{(Color online) (a) $\Delta F$ as a function of magnetic field for \textbf{H}$\parallel$\textbf{ab} at selected temperatures. Solid lines are 60 T short-pulse measurements and
symbols (triangle and square) are 92 T multi-shot measurements. Open circles are $\Delta F$ taken at 28 K as a normal-state background signal. The determined $H_{c2}$ is indicated by
arrows. The dashed lines represent the linear extrapolation of $\Delta F$ to the normal state. The horizontal dash-dotted line is guide to eye. The inset shows the temperature
dependence of $\Delta F$ at $H$ = 0 for three samples. (b) $\Delta F$ as a function of magnetic field for \textbf{H}$\parallel$\textbf{c} at selected temperatures. Due to the small
size of the sample used, the $\Delta F$ measured in the multi-shot magnet is small compared to 60 T curve. Inset shows $\Delta F$ curves at $T$ = 4 K from both magnet systems. The
arrow indicates the $H_{c2}$ determined by 92 T measurement.}
\label{Fig2}%
\end{figure}

In zero field the superconducting transition temperature, determined from the d($\Delta F$)/d$T$ analysis, gives $T_{c}$ = 26.5 K for three samples, used in the both short pulse 60 T
and multi-shot 95 T magnet (inset in Fig. \ref{Fig2} (a)); as the temperature decreases, $\Delta F$ sharply increases at $T_{c}$. The evolution of superconductivity with magnetic
field is shown in Fig. \ref{Fig2} (a) for \textbf{H}$\parallel$\textbf{ab} and Fig. \ref{Fig2} (b) for \textbf{H}$\parallel$\textbf{c}. At this point it should be noted that the phase
transitions determined by the rf technique are consistent with off-set criteria of resistance measurements shown in Ref. \cite{Mun2011}. In comparison with the samples reported in
Ref. \cite{Ni2011}, samples studied in this work show a slightly higher $T_{c}$, related to the different Pt doping concentration in the Fe$_{2}$As$_{2}$ layer.

\begin{figure}
\centering
\includegraphics[width=0.9\linewidth]{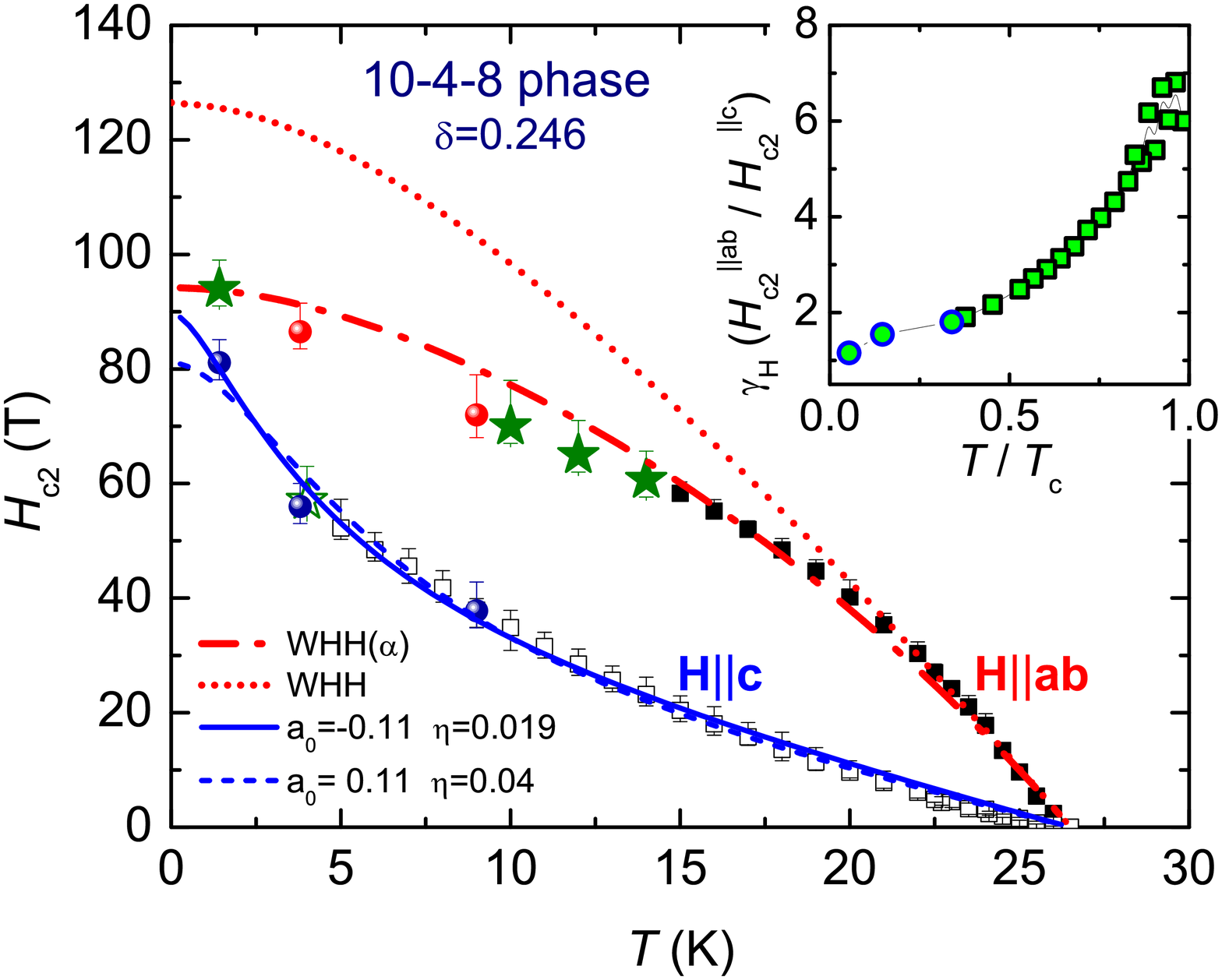}
\caption{(Color online) Anisotropic $H_{c2}(T)$ for 10-4-8 phase single crystals. The square symbols are obtained from the 60 T short pulse and circles are taken from the 92 T
multi-shot magnet. The star symbols are extracted by the linear extrapolation of $\Delta F$, as shown by dashed lines in Figs. \ref{Fig2} (a) and (b). For
\textbf{H}$\parallel$\textbf{ab}, the dotted line are the WHH $H_{c2}$ and the dash-dotted line represents the fit with $H_{c2}^{*}$ = $H_{c2}/\sqrt{1+\alpha^2}$ ($\alpha \approx$
0.9). For \textbf{H}$\parallel$\textbf{c}, the solid line is a best fit of the two-band model to the experimental data with $a_{0}$ = -0.11 and $\eta$ = 0.019 and the dashed line
represents the fit with a fixed value of $a_{0}$ = 0.11. Inset shows the temperature dependence of the anisotropy $\gamma_{H} \equiv
H_{c2}^{\parallel\textbf{ab}}/H_{c2}^{\parallel\textbf{c}}$. Squares and circles are determined from $H_{c2}$ taken in the 60 T short pulse and 95 T multi-shot magnet, respectively.}
\label{Fig3}%
\end{figure}

Using the deviation from the normal state criterion, discussed in Ref. \cite{Mun2011}, the $H_{c2}(T)$ curve is inferred from the $\Delta F$ vs $H$ plots shown in Figs. \ref{Fig2} (a)
and (b); the determined $H_{c2}$ is indicated by arrows. Note that the $H_{c2}$ error bar is estimated by the difference between the $H_{c2}$ values determined by the point deviating
from the background and by taking the slope of the rf signal intercepting the slope of the normal state background \cite{Mun2011}. In addition, the hysteresis detected in 92 T
multi-shot measurements (see Fig. \ref{Fig1}) is considered to determine the error bar. The obtained $H_{c2}(T)$ curves for both \textbf{H}$\parallel$\textbf{ab} and
\textbf{H}$\parallel$\textbf{c} are plotted in Fig. \ref{Fig3}, as determined from the $H$ $<$ 60 T in a short pulse (squares) and from the $H$ $<$ 92 T in a multi-shot (circles)
taken from the down sweep of the magnetic field. The star symbols in Fig. \ref{Fig3} are extracted by linear extrapolation of $\Delta F$ curves at $H$ $<$ 60 T and $H$ $<$ 92 T, as
shown by dashed lines in Figs. 2 (a) and (b). As a representative curve, $\Delta F$ taken at 4 K in a 60 T short pulse magnet is presented in the inset of Fig. \ref{Fig2} (b), where
the critical field determined by extrapolating $\Delta F$ curve is consistent with the one determined by down sweep of the 92 T magnetic field. The shapes of the $H_{c2}(T)$ curves
for \textbf{H}$\parallel$\textbf{ab} and \textbf{H}$\parallel$\textbf{c} in Fig. \ref{Fig3} clearly show the different temperature dependence. A conventional linear field dependence
of $H_{c2}$ is observed close to the $T_{c}$, with clearly different slopes for the two field orientations. In the low field region, the $H_{c2}$ curves are consistent with earlier
resistance study \cite{Ni2011}. Toward higher fields, $H_{c2}^{\parallel\textbf{ab}}$ presents a tendency to saturate, whereas the curve for $H_{c2}^{\parallel\textbf{c}}$ shows a
steep increase.

The anisotropy coefficient $\gamma_{H}$ is plotted in the inset of Fig. \ref{Fig3}. One can see that the value of $\gamma_{H}$ is $\sim$7 near $T_{c}$ and then decreases considerably
to $\sim$1 with decreasing temperature. This temperature dependence of $\gamma_{H}$ is similar to other Fe-based superconductors. Close to $T_{c}$ the $\gamma_{H}$ of the 10-4-8 phase
is bigger than that of the 122-type \cite{Altarawneh2008, Ni2008, Kano2009} and is comparable to the 1111-type \cite{Hunte2008, Jaroszynski2008}. Based on the results measured here
and in other studies, this temperature dependence of $\gamma_{H}$ and a weak anisotropy at $T \ll T_{c}$ seem to be a general feature of the Fe-based superconductors. The strong
temperature dependence of $\gamma_{H}$ can be a signature of multi-band superconductivity. Based on the recent literature \cite{Kogan2011}, however, the $\gamma_{H}$ can even have a
weak temperature dependence for a single band $d$-wave order parameter. Thus, our $\gamma_{H}$ results need to be carefully analyzed.

The zero temperature limit of $H_{c2}$ can be predicted by using the Werthamer-Helfand-Hohenberg (WHH) theory \cite{WHH1966}, which gives $H_{c2}$ =
0.69$T_{c}$(d$H_{c2}$/d$T$)$\mid_{T_{c}}$. The values of $H_{c2}$(0) for \textbf{H}$\parallel$\textbf{ab} and \textbf{H}$\perp$\textbf{c} are estimated to be
$H_{c2}^{\parallel\textbf{ab}}$ $\sim$ 132 T and $H_{c2}^{\parallel\textbf{c}}$ $\sim$ 24 T, respectively. For $T$ $<$ 0.6 $T_{c}$, the $H_{c2}^{\parallel\textbf{ab}}$ already exceeds
the Pauli limit ($H_{p}$ = 1.84 $T_{c}$) \cite{Clogston1962}, giving $H_{p}$ $\sim$ 48.7 T. Clearly, these values do not capture the salient physics of this compound.

A number of experimental and theoretical papers have discussed the Fe-based superconductors as a two-band system \cite{Hunte2008, Jaroszynski2008, Kano2009, Lee2008, Eschrig2009,
Weyeneth2009, Paglione2010}. Certainly, multi-band superconductivity can be manifested in the $H_{c2}(T)$ curve, as shown in MgB$_{2}$ \cite{Gurevich2003}. The clear upward curvature
in $H_{c2}^{\parallel\textbf{c}}$ in 10-4-8 phase cannot be explained by the one-gap WHH theory and, therefore, suggests the multi-band nature of superconductivity. In the dirty
limit, the two-band model for $H_{c2}$ \cite{Gurevich2003}, which takes into account both orbital and Zeeman pair breaking for negligible interband scattering, is thus used to fit the
data so that a qualitative understanding can be reached. Assuming negligible interband scattering, the equation of $H_{c2}(T)$ can be written as $a_{0}[lnt + U(h/t)] [lnt + U(\eta
h/t)] + a_{2}[lnt+U(\eta h/t)] + a_{1}[lnt + U(h/t)] = 0$, where $a_{0} = 2(\lambda_{11}\lambda_{22}-\lambda_{12}\lambda_{21})$, $a_{1} = 1 + (\lambda_{11}-\lambda_{22})/\lambda_{0}$,
$a_{2} = 1 - (\lambda_{11}-\lambda_{22})/\lambda_{0}$, $\lambda_{0} = [(\lambda_{11}-\lambda_{22})^2+4\lambda_{12}\lambda_{21}]^{1/2}$, $h = H_{c2}D_1/2\phi_0 T$, $t = T/T_c$, $\eta =
D_{2}/D_{1}$, and $U(x) = \psi(x+1/2)-\psi(x)$. $\psi(x)$ is the di-gamma function, $\lambda_{11}$ and $\lambda_{22}$ are the intraband BCS coupling constants, while $\lambda_{12}$
and $\lambda_{21}$ are the interband BCS coupling constants, $D_{1}$ and $D_{2}$ are the in-plane diffusivity of each band.

As can be inferred from the equation, if $a_{0}$ $>$ 0, the intraband coupling dominates while if $a_{0}$ $<$ 0, the interband coupling dominates. We assume $\lambda_{11}$ =
$\lambda_{22}$, such that $a_{1}$ = $a_{2}$ = 1, so that the shape of the curve is only determined by fitting parameters $a_{0}$ and $\eta$. The best fit
$H_{c2}^{\parallel\textbf{c}}$ curve with $a_{0}$ = -0.11 and $\eta$ = 0.019 is shown as a solid blue line in Fig. \ref{Fig3}, and can be seen to agree very well with the experimental
data. Although $a_{0}$ $<$ 0 for the best fit, it cannot be concluded that the interband coupling dominates. The dashed blue curve, obtained by a fixed $a_{0}$ = 0.11 (intraband
coupling dominates), also fits the data if the error bars are considered. The small $\eta$ implies stronger scattering in one of the bands. Although the exact determination of the band
coupling is not possible from our data analysis, the $H_{c2}$ curvature for $H_{c2}^{\parallel\textbf{c}}$ can be explained well by the two-band model.

For $H_{c2}^{\parallel\textbf{c}}$, the orbital limit is comparable to the Pauli limit and the two-band model works well. For $H_{c2}^{\parallel\textbf{ab}}$, however, the two-band
model fit does not result in physically meaningful parameters since the orbital limit is much larger than the Pauli limit. When including Pauli paramagnetism, the
$H_{c2}^{\parallel\textbf{ab}}(T)$ curve can be fit with $H_{c2}^{*}$ = $H_{c2}/\sqrt{1+\alpha^2}$, where the Maki parameter $\alpha$ = $\sqrt{2} H_{c2}/H_{P}$ \cite{Maki1966}. A
rather good agreement between the data and fitting is observed as shown by the dash-dotted red line in Fig. \ref{Fig3}.

An interesting behavior in the $H_{c2}$ curves for 10-4-8 phase is that the \textbf{H}$\parallel$\textbf{ab} and \textbf{H}$\parallel$\textbf{c} curves as seen to merge at low
temperatures ($\gamma_{H}$ $\sim$ 1). For optimally Co-doped BaFe$_{2}$As$_{2}$ \cite{Kano2009}, the $H_{c2}$ curves for the two field orientations also appear to meet at $T$ = 0 K.
The Te-doped FeSe system \cite{Khim2010} show crossing $H_{c2}$ curves between \textbf{H}$\parallel$\textbf{ab} and \textbf{H}$\parallel$\textbf{c}, because of the subsequent
flattening of the $H_{c2}^{\parallel\textbf{ab}}$ curve at low temperatures. However, the $H_{c2}$ curves in Co-doped BaFe$_{2}$As$_{2}$ \cite{Ni2008}, K$_{0.8}$Fe$_{1.76}$Se$_{2}$
\cite{Mun2011}, LiFeAs \cite{Cho2011}, and oxypnictide systems \cite{Hunte2008, Jaroszynski2008, Lee2008} do not show such a crossing behavior.

In summary, we have determined the upper critical fields $H_{c2}(T)$ of the Ca$_{10}$(Pt$_{4-\delta}$As$_{8}$)((Fe$_{0.97}$Pt$_{0.03}$)$_{2}$As$_{2}$)$_{5}$ ($\delta$ $\approx$ 0.246)
superconductor by measuring the rf contactless penetration depth up to 92 T in pulsed magnetic fields, which enable access to the complete evolution of $H_{c2}(T)$ curves. It was
found that this supercondcutor shows very high $H_{c2}(0)$ values ($\sim$90 T for \textbf{H}$\parallel$\textbf{c} and $\sim$94 T for \textbf{H}$\parallel$\textbf{ab}) and relatively
weak anisotropy of superconductivity at low temperatures. For \textbf{H}$\parallel$\textbf{ab}, $H_{c2}(T)$ can be described by taking into account Pauli paramagnetism, whereas for
\textbf{H}$\parallel$\textbf{c}, $H_{c2}(T)$ exhibits a steep increase at low temperatures, quite possibly due to the multi-band effect.

\begin{acknowledgments}
The authors thank M. Gordon, D. Roybal, Y. Coulter, J. Betts, M. Pacheco and C. Swenson for 100 T technical support; E. Mun thanks H. Park for useful discussions. Work at the National
High Magnetic Field Laboratory was supported by the US National Science Foundation through Cooperative Grant No. DMR901624, the State of Florida, and the US Department of Energy, and
US office of Science project "Science at 100 T". This work was supported by the AFOSR MURI on superconductivity.

\end{acknowledgments}

\end{document}